# Femtosecond X-ray Laser induced transient electronic phase change observed in fullerene $C_{60}$

Brian Abbey[1,2], Ruben A. Dilanian[2], Connie Darmanin[2,3], Rebecca A. Ryan[2], Corey T. Putkunz[2], Andrew V. Martin[2], Victor Streltsov[3], Michael W. M. Jones[1], Naylyn Gaffney[4], Felix Hofmann[5], Garth J. Williams[6], Sébastian Boutet[6], Marc Messerschmidt[6], M. Marvin Siebert[6], Sophie Williams[2], Evan Curwood[2], Eugeniu Balaur[1], Andrew G. Peele[1,7], Keith A. Nugent[2] and Harry M. Quiney[2]*

[1]ARC Centre of Excellence for Coherent X-Ray Science , Department of Physics, La Trobe University, Bundoora, Victoria 3086, Australia
[2]ARC Centre of Excellence for Coherent X-Ray Science, School of Physics, The University of Melbourne, Victoria, 3010, Australia
[3]ARC Centre of Excellence for Coherent X-Ray Science, CSIRO Molecular and Health Technologies and Preventative Health Flagship, 343 Royal Parade, Parkville, Vic. 3052, Australia.
[4]ARC Centre of Excellence for Coherent X-Ray Science, Centre for Atom Optics and Ultrafast Spectroscopy, Swinburne University of Technology, Melbourne 3122, Australia
[5]Chemistry Department, Massachusetts Institute of Technology, 77 Massachusetts Avenue, Cambridge, MA 02139-4307, USA
[6]SLAC National Accelerator Laboratory, 2575 Sand Hill Road, Menlo Park, CA 94025, USA
[7]Australian Synchrotron, 800 Blackburn Rd, Clayton 3168, Australia
*Email: quiney@unimelb.edu.au

X-ray Free-Electron Lasers (XFELs) [1] deliver X-ray pulses with a coherent flux that is approximately eight orders of magnitude greater than that available from a modern third generation synchrotron source. The power density in an XFEL pulse may be so high that it can modify the electronic properties of a sample on a femtosecond timescale. Exploration of the interaction of intense coherent X-ray pulses and matter is of both intrinsic scientific interest, and of critical importance to the interpretation of experiments that probe the structures of materials using high-brightness femtosecond XFEL pulses. In this letter, we report observations of the diffraction of an extremely intense 32 fs nanofocused X-ray pulses by a powder sample of crystalline $C_{60}$. We find that the diffraction pattern at the highest available incident power exhibits significant structural signatures that are absent in data obtained at both third-generation synchrotron sources **or from XFEL sources operating at low output power. These signatures are consistent with a highly ordered structure that does not correspond with any previously known phase of crystalline $C_{60}$. We argue that these data indicate the formation of a transient phase that is formed by a dynamic electronic distortion induced by inner-shell ionisation of at least one carbon atom in each $C_{60}$ molecule.**

To investigate the mechanisms of electronic damage during femtosecond X-ray diffraction measurements we devised an experiment in which a powdered sample of crystalline $C_{60}$ was subjected to XFEL pulses of varying intensity [2]. The experiment employed a 10 keV (1.24 Å wavelength) X-ray beam produced at the Linac Coherent Light Source (LCLS) [1], focused by a Kirkpatrick-Baez (KB) mirror system to an area of less than 200 nm² full-width at half-maximum (FWHM) with a per-pulse fluence of 1.5 x $10^{12}$ photons at peak power. The sample was a fixed target of $C_{60}$ nanocrystals, each of which was comparable in size to the beam demonstrated by the fact that each XFEL shot produced a collection of Bragg spots from a small number of differently oriented crystals. The sample was produced by crushing 99.9+% pure, macroscopic $C_{60}$ powder (purchased from SES research [3]). A layer thin enough to be within the ~100 μm depth of focus produced by the KB system was supported by a 10 μm thick kapton polymide backing. At this energy the scattered kapton background (which primarily occurs at low-q) can readily be subtracted from the $C_{60}$ data. The sample was scanned continuously in the beam at a pulse repetition rate of 1 Hz with pulses in the range of 27-36 fs FWHM based on the electron bunch duration. Attenuation upstream of the sample (Fig. 1) allowed the incident intensity to vary between 10% and 100% of the maximum value, permitting the introduction of intensity variations without changing samples or disrupting the vacuum conditions.

Forward-scattered diffraction data (Fig. 2a-2c) were collected at an estimated peak power density of 7 x $10^{20}$ W cm² when unattenuated, using the Cornell-SLAC Pixel Array Detector (CSPAD) [4] containing 1516 x 1516 square pixels each of area 110 μm², placed 79 mm downstream of the plane of the sample. Fig. 2b shows an expansion of data collected using 10% power and 2c an expansion of the same region with 100% power. The direct beam was allowed to pass through a gap in the centre of the CSPAD detector modules. Each shot involved diffraction from a small number of nanocrystals per shot, producing discrete Bragg peaks in each frame of data. Bragg-diffraction data more than 1000 single shot measurements was summed to produce a complete powder pattern for both 10% and 100% data (Fig. 2d and 2e) with a $C_{60}$ powder outer ring resolution of better than 2 Å. Synchrotron microdiffraction data was also collected at the Australian Synchrotron (AS). The XFEL data and the data collected at the AS used the same sample of $C_{60}$.

A significant dependence of the diffraction pattern on incident intensity is observed for diffraction angles, $2\theta \geq 20°$, while the diffraction pattern at lower resolution is comparatively insensitive to the incident intensity. Structural analysis shows that the XFEL data at 10% power and AS powder diffraction patterns correspond to the well-known room-temperature face centred cubic (FCC) structure of $C_{60}$ [5], twinned along the [111] direction of the crystal lattice. The twined FCC structure is defined by the stacking sequence of close-packed (111) planes; the presence of such stacking faults leads to the displacement of Bragg reflections from their ideal positions [6]. The directions of such displacements are dependent on the identity of the reflection.

The dramatic dependence of the Bragg reflection structure on the incident XFEL pulse intensity cannot, however, be explained within the framework of the standard FCC structure. Comparison of the powder patterns for the synchrotron, attenuated, and full intensity data (Fig. 2) show a clear suppression of the intensity of some peaks and a marked increase in others for the full intensity data above a resolution of 3.5 Å ($2\theta \geq 20°$). These observations provide direct evidence that the scattering factors of the $C_{60}$ molecules have been modified by the XFEL pulse in a manner that preserves the translational symmetry of the twinned $C_{60}$ crystal. The data suggests a new, albeit transient, structural phase of the $C_{60}$ crystal has been created through interaction with the nano-focused XFEL beam.

The room temperature diffraction pattern of crystalline $C_{60}$ indicates that there is no preferred orientation of the molecules in the FCC lattice because of rapid molecular rotation. The effective molecular form factor (MFF) of $C_{60}$ is designated $\Phi(\mathbf{q})$ and is assumed to be centro-symmetric at room temperature. There are four molecules in the FCC unit cell of $C_{60}$ located at the normalized coordinates $(0,0,0), (1/2,1/2,0), (1/2,0,1/2)$ and $(0,1/2,1/2)$. The scattering amplitude, $F(\mathbf{q})$, of the Bragg reflection with Miller indices *(h,k,l)* can be written in the forms $F_A(\mathbf{q}) = \Phi_1(\mathbf{q}) + \Phi_2(\mathbf{q}) + \Phi_3(\mathbf{q}) + \Phi_4(\mathbf{q})$ if $h+k=2n$, $h+l=2n$, $k+l=2n$, or $F_B(\mathbf{q}) = (\Phi_1(\mathbf{q}) \pm \Phi_2(\mathbf{q})) \mp (\Phi_3(\mathbf{q}) \pm \Phi_4(\mathbf{q}))$ otherwise. For FCC structures $F_B(\mathbf{q}) = 0$, so these Bragg reflections are designated 'forbidden'. The XFEL data collected at 10% power and the data collected at AS are essentially identical and satisfy these selection rules on forbidden reflections in twinned FCC structures comprised of centro-symmetric molecules. The data collected at 100% power using XFEL pulses, however, violates these selection rules and does not correspond to the low-temperature structure of $C_{60}$, which involves orientational alignment of molecules in the lattice.

In order to explain the observed diffraction of $C_{60}$ at the highest illuminated intensity, we seek to identify a process by a coherent redistribution of electron density on a femtosecond timescale, rather than a classical change in structural phase of either the lattice parameters or the individual $C_{60}$ molecules, which is presumed to occur on a much longer timescale. Since the Auger lifetime is ~10 fs, the majority of holes formed via ionisation will be core-shell. Inner-shell ionisation of the $C_{60}$ molecules modifies $\Phi(\mathbf{q})$ and can modify the intensities of forbidden reflections if the modification reduces the local molecular symmetry while preserving the space-group symmetry of the lattice. Further details of the crystallographic analysis of the data can be found in the Methods section.

The random allocation of localized ionization events on $C_{60}$ molecules simply contributes a diffuse background signal similar to the orientational ordering effect. Such a stochastic atomic process is not, however, consistent with the data in Fig. 2, which indicate a strong, and intrinsically coherent effect, in the 100% intensity XFEL data. If, on the other hand, we consider the possibility that the creation of inner-shell vacancies in a regular array of $C_{60}$ molecules is a process coupled by Coulomb interactions, the distribution of the holes created by X-ray photoionization can evolve within the decoherence time of this process, which is typically fixed by the Auger lifetime to be around 5-10 fs. Delocalized inner-shell vacancies may evolve in time to produce a spatially regular array of $C_{60}^+$ ions whose local symmetry is reduced from spherical to cylindrical (depicted in Fig. 3). This leads to an additional phase contribution to the MFF of $C_{60}^+$ ions that invalidates the selection rules for forbidden reflections that dominates the diffraction patterns of the data from AS and the XFEL data obtained using 10% incident power. If, on average, one inner shell electron is removed from each $C_{60}$ molecule during a period that is short (~1 fs) compared to the decoherence time (~10 fs), the inner-shell vacancies can evolve into a dipolar charge distribution that preserves the long-range translational symmetry of the unperturbed crystalline lattice. One also expects to find some diffuse scattering due to fluctuations in the hole densities and their imperfect alignments along the cylindrical symmetry axis; this background scattering is readily isolated from the Bragg peaks.

We constructed a forward model of a $C_{60}$ crystalline diffraction pattern modified by interaction with the XFEL pulse at 100% intensity by including both twinning of the FCC lattice and collective alignment of ionized molecules. In our analysis we assumed that ionized molecules are aligned along one of the main diagonals of the cubic lattice. The ion located at $(0,0,0)$ was aligned along the $[111]$ direction, the ion at $(1/2,1/2,0)$ along the $[\bar{1}11]$ direction, the ion at $(1/2,0,1/2)$ along the $[\bar{1}\bar{1}\bar{1}]$ direction and the ion at $(0,1/2,1/2)$ along the $[\bar{1}\bar{1}\bar{1}]$ direction. Figure 2(f) shows the effect of twining of the FCC lattice which provides a match to the 10%

XFEL and AS data. Adding the effect of the collective alignment of the non-centro-symmetrical $C_{60}$ molecule-ions to give a combined contribution from both effects provides a good match to the 100% XFEL data. Our model of the electronic transformation in the $C_{60}$ crystal requires neither a change in lattice symmetry and parameters nor a relocation of the atomic positions. This is consistent with the underlying assumption that significant nuclear motion should not occur on the 40 fs timescale over which diffraction information is collected.

In our model, we require a sufficient level of ionization to match the experimental data. In the 100% intensity case, we assume a 200 x 200 nm probe with $10^{12}$ photons indicating that there are of the order of 10 ionizations per molecule during the interaction with the 40 fs pulse. This estimate is sufficient, however, to suggest that significant inner-shell photoionization can occur on a timescale shorter than the Auger lifetime of ~10 fs and that the induced inner-shell hole distribution has sufficient time to evolve into a periodic structure under the influence of their mutual Coulomb interactions.

The time-scale for the realignment of the dipoles, which is depicted in Fig. 3, is related to the strength of the Coulomb interaction between the induced charges on neighbouring $C_{60}$ molecules. The diameter of a $C_{60}$ molecule is 7 Å, which we can take as both the typical distance between charges and also the distance a single charge needs to be displaced in order for a dipole to realign. The strength of the Coulomb potential induced by a single hole state at distance of 7 Å is on the order of 1 eV per elementary charge, which is sufficient to displace a free electron a few Angstroms within a femtosecond duration. Although a time-dependent quantum mechanical model of the hole dynamics is more complex, this simple estimate indicates that the time scale of hole realignment is of the right order for our physical model to be plausible.

In conclusion, we have observed a new transient phase of $C_{60}$ which exists during the interaction of standard fullerene $C_{60}$ with a highly focused, sub-40 fs, XFEL pulse. Surprisingly, the ionisation of a significant fraction of the atoms in the $C_{60}$ crystals does not, as one might expect, lead to disorder. In fact, we suggest here that it induces a different kind of ordering of the electronic structure, resulting in a change in the crystalline diffraction pattern which is *not* mediated by nuclear motion. This results not in a subtle change in the diffraction data but rather to a wholesale modification that is the characteristic signature of a quantum mechanical coherence effect. Our interpretation of the results suggests that the effects reported here require a high degree of molecular symmetry. However the model proposed here needs further experimental testing and it is by no means certain that similar femtosecond electronic re-arrangement will be absent from structures consisting of lower symmetry molecules. These are open questions which can only be answered via further experiments. What is clear is that to understand the conditions produced by highly focused XFEL beams requires that we move beyond classical electrostatic X-ray diffraction theory, taking into account the electrodynamic changes induced by the measurement process itself.

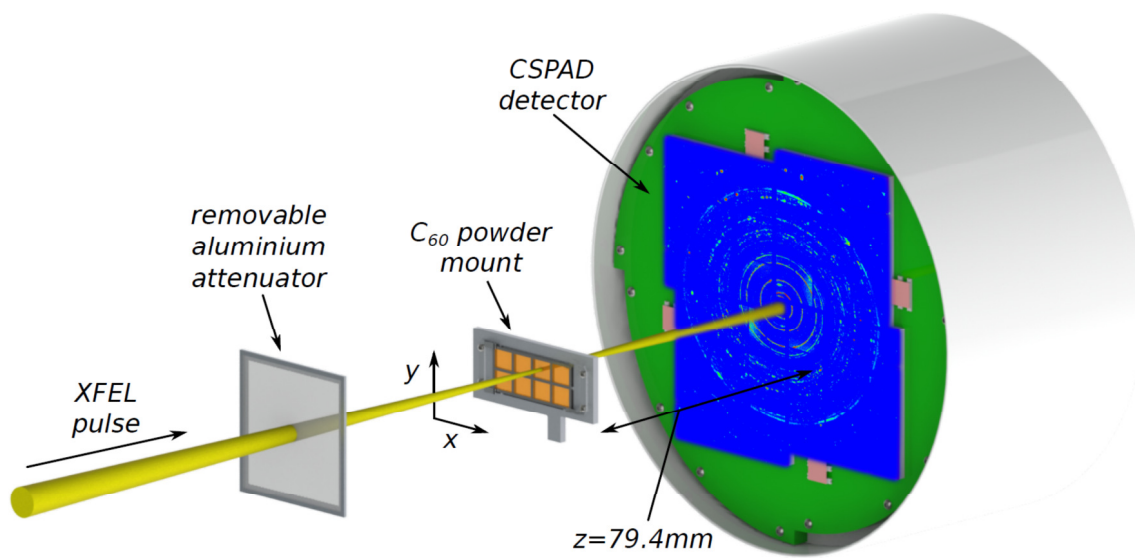

Figure 1: Schematic of the experimental geometry used to collect the diffraction data. 10 keV 30 fs XFEL pulses provided simultaneous pump and probe functions when passing through the $C_{60}$ sample, which was suspended on kapton. Upstream aluminium attenuators were inserted to provide 10% incident beam power. Similar attenuators were placed downstream of the sample (not displayed) to inhibit saturation of the CXI CSPAD detector position. The intense direct beam pased through a small hole in the CSPAD detector modules.

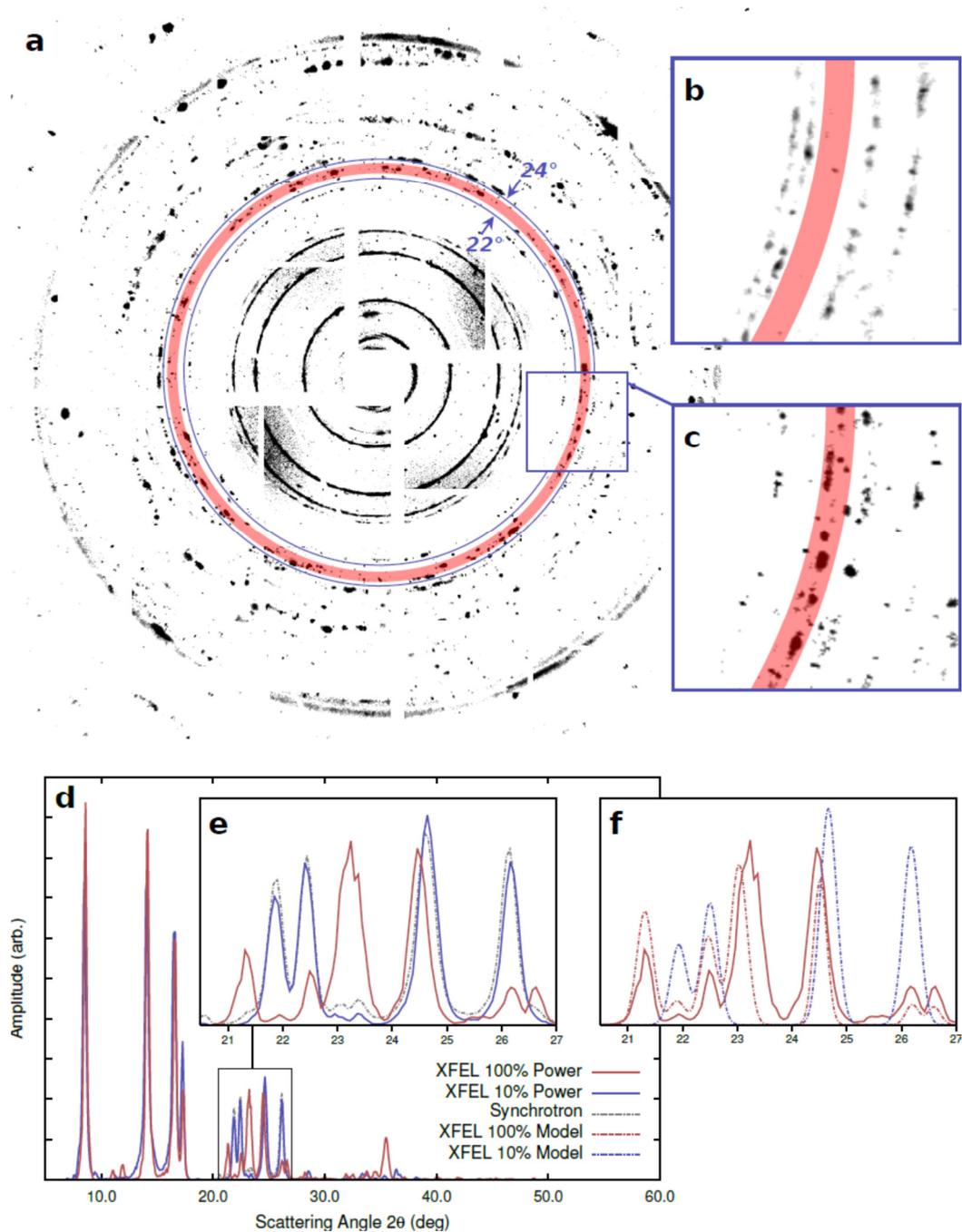

Figure 2: a) Summed diffraction data from 2500 single shots recorded at 100% power. Visible Bragg peaks indicate the presence of sufficiently large crystals within the XFEL illumination. (b) Zoomed region showing Bragg peaks at 10% power at positions consistent with room temperature FCC structure. (c) The same region as in (b) collected at 100% power. To enhance contrast the $\sqrt{(I)}$ has been displayed. d) – f) show radially averaged diffraction data. The blue curve shows the $C_{60}$ powder diffraction pattern created via summing 1000 individual shots (containing single crystal and partial powder data) at 10% power, in agreement with the structure of $C_{60}$ at room temperature recorded during synchrotron experiments (black curve). The red curve shows the powder diffraction pattern from 2500 individual shots at 100% power. The significant variation in peaks at approximately 0.3 inverse angstroms (20-30 degrees) represents a phase change of the molecular lattice. The red bar indicates the same reciprocal space region as in a)- c).

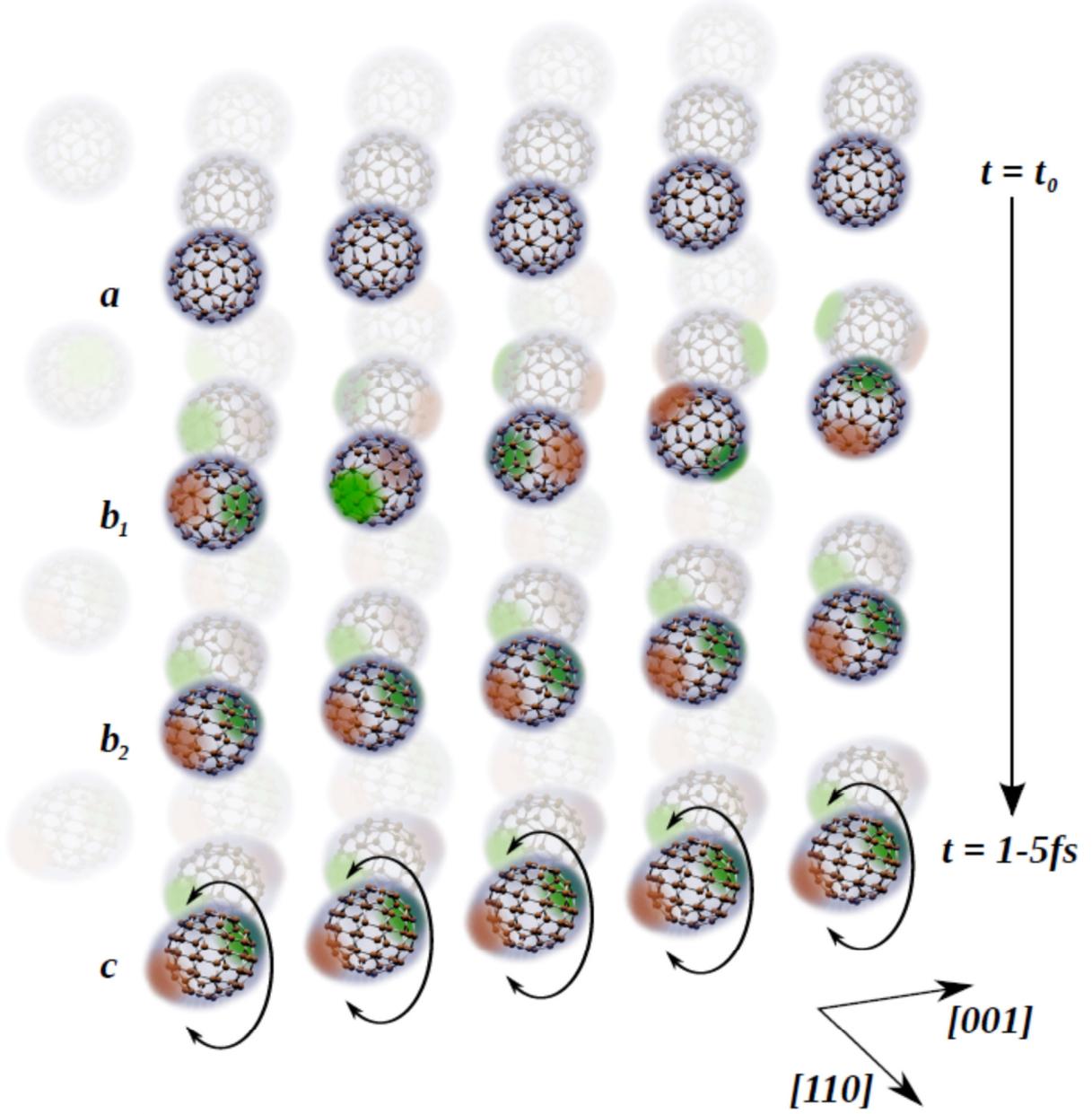

Figure 3: The initial interaction of the XFEL pulse with the unperturbed (a) $C_{60}$ crystal. Ionisation (b1) and alignment (b2) of the molecules is followed by an axially aligned distortion (c) of the electronic charge distribution.

**METHODS**

To explain why the existence of this new crystalline phase for $C_{60}$ depends on the XFEL intensity, let us consider the amplitude of X-rays scattered from the $C_{60}$ molecules packed in the FCC lattice defined as:

$$F(\mathbf{q}) = \sum_{k=1}^{4} \Phi_k(\mathbf{q}) \exp(i 2\pi \mathbf{q} \cdot \mathbf{R}_k). \qquad (1)$$

Here, $\mathbf{q}$ is the scattering vector, $\mathbf{R}_k$ is the position of the $k^{\text{th}}$ C$_{60}$ molecule in the unit cell and $\Phi_k(\mathbf{q})$ define the scattering from the $k^{\text{th}}$ C$_{60}$ molecule, the molecular form factor (MFF). Given four positions, $\mathbf{R}_k$, of the C$_{60}$ molecules in the FCC lattice, the scattering amplitude of the Bragg reflection with the Miller indexes (*hkl*) can be rewritten in the form $F_A(\mathbf{q}) = \Phi_1(\mathbf{q}) + \Phi_2(\mathbf{q}) + \Phi_3(\mathbf{q}) + \Phi_4(\mathbf{q})$ if $h+k=2n$, $h+l=2n$, $k+l=2n$, and $F_B(\mathbf{q}) = (\Phi_1(\mathbf{q}) \pm \Phi_2(\mathbf{q})) \mp (\Phi_3(\mathbf{q}) \pm \Phi_4(\mathbf{q}))$ otherwise. Since at room temperature C$_{60}$ molecules are randomly oriented in the crystal lattice, one can model the C$_{60}$ molecule as a sphere whose surface is randomly covered by carbon atoms. In this case, reflections that satisfy conditions for $F_B(\mathbf{q})$ are 'forbidden' ($F_B(\mathbf{q})=0$). When the orientations of molecules are fixed [6], the spherical approximation is incorrect and Bragg reflections that satisfy conditions for $F_B(\mathbf{q})$ will be observed in the diffraction pattern. It can be shown, however, that since the ideal C$_{60}$ molecule is centrosymmetric and, therefore, the MFFs are real functions, $|F_A(\mathbf{q})| \geq |F_B(\mathbf{q})|$ for all reflections below the 2.5 A resolution. Consequently, the differences between the 10% and 100% intensity XFEL data in the range $20^\circ \geq 2\theta \geq 27^\circ$, Fig. 2e, cannot be described by the specific orientational ordering of the C$_{60}$ molecules in the ideal crystal lattice.

Meanwhile, ionisation of the carbon atoms in the C$_{60}$ molecules can modify the $|F_A(\mathbf{q})|/|F_B(\mathbf{q})|$ ratio significantly. The ionisation induces a dipole moment in each C$_{60}$ molecule, which changes the resulting symmetry of the C$_{60}$ molecule. The loss of the centrosymmetry leads to an additional phase contribution in the scattering amplitude, since the MFFs of C$_{60}$ molecules are no longer real but complex functions. Moreover, neighbouring C$_{60}$ dipoles could align via a Coulomb interaction on a timescale significantly shorter than the XFEL pulse. Such correlated effect will change the resulting intensities, $I_A(\mathbf{q})$ and $I_B(\mathbf{q})$ of Bragg reflections.